\documentclass[pre,twocolumn,amssymb]{revtex4}

\usepackage{graphicx}
\usepackage{color}
\usepackage{epsfig}
\usepackage{latexsym}
\usepackage{bm}
\usepackage{ulem}

\begin{document}
\renewcommand{\ni}{{\noindent}}
\newcommand{\dprime}{{\prime\prime}}
\newcommand{\be}{\begin{equation}}
\newcommand{\ee}{\end{equation}}
\newcommand{\bea}{\begin{eqnarray}}
\newcommand{\eea}{\end{eqnarray}}
\newcommand{\nn}{\nonumber}
\newcommand{\bk}{{\bf k}}
\newcommand{\bQ}{{\bf Q}}
\newcommand{\q}{{\bf q}}
\newcommand{\s}{{\bf s}}
\newcommand{\bN}{{\bf \nabla}}
\newcommand{\bA}{{\bf A}}
\newcommand{\bE}{{\bf E}}
\newcommand{\bj}{{\bf j}}
\newcommand{\bJ}{{\bf J}}
\newcommand{\bs}{{\bf v}_s}
\newcommand{\bn}{{\bf v}_n}
\newcommand{\bv}{{\bf v}}
\newcommand{\la}{\langle}
\newcommand{\ra}{\rangle}
\newcommand{\dg}{\dagger}
\newcommand{\br}{{\bf{r}}}
\newcommand{\brp}{{\bf{r}^\prime}}
\newcommand{\bq}{{\bf{q}}}
\newcommand{\hx}{\hat{\bf x}}
\newcommand{\hy}{\hat{\bf y}}
\newcommand{\bS}{{\bf S}}
\newcommand{\cU}{{\cal U}}
\newcommand{\cD}{{\cal D}}
\newcommand{\bR}{{\bf R}}
\newcommand{\pll}{\parallel}
\newcommand{\sumr}{\sum_{\vr}}
\newcommand{\cP}{{\cal P}}
\newcommand{\cQ}{{\cal Q}}
\newcommand{\cS}{{\cal S}}
\newcommand{\ua}{\uparrow}
\newcommand{\da}{\downarrow}
\newcommand{\red}{\textcolor {red}}
\newcommand{\blu}{\textcolor {blue}}

\def\lsim {\protect \raisebox{-0.75ex}[-1.5ex]{$\;\stackrel{<}{\sim}\;$}}
\def\gsim {\protect \raisebox{-0.75ex}[-1.5ex]{$\;\stackrel{>}{\sim}\;$}}
\def\lsimeq {\protect \raisebox{-0.75ex}[-1.5ex]{$\;\stackrel{<}{\simeq}\;$}}
\def\gsimeq {\protect \raisebox{-0.75ex}[-1.5ex]{$\;\stackrel{>}{\simeq}\;$}}


\title{Hydrodynamics, density fluctuations and universality in conserved stochastic sandpiles }

\author{Sayani Chatterjee, Arghya Das, and Punyabrata Pradhan}

\affiliation{Department of Theoretical Sciences, S. N. Bose National Centre for Basic Sciences, Block - JD, Sector - III, Salt Lake, Kolkata 700106, India }

\begin{abstract}

\noindent{We study conserved stochastic sandpiles (CSSs), which exhibit an active-absorbing phase transition upon tuning density $\rho$. We demonstrate that a broad class of CSSs possesses a remarkable  hydrodynamic structure: There is an Einstein relation $\sigma^2(\rho) = \chi(\rho)/D(\rho)$, which connects bulk-diffusion coefficient $D(\rho)$, conductivity $\chi(\rho)$ and mass-fluctuation, or scaled variance of subsystem mass, $\sigma^2(\rho)$. Consequently, density large-deviations are governed by an equilibriumlike chemical potential $\mu(\rho) \sim \ln a(\rho)$ where $a(\rho)$ is the activity in the system. Using the above hydrodynamics, we derive two scaling relations: As $\Delta = (\rho - \rho_c) \rightarrow 0^+$, $\rho_c$ being the critical density, (i) the mass-fluctuation $\sigma^2(\rho) \sim \Delta^{1-\delta}$ with $\delta=0$ and (ii) the dynamical exponent $z = 2 + (\beta -1)/\nu_{\perp}$, expressed in terms of two static exponents $\beta$ and $\nu_{\perp}$ for activity $a(\rho) \sim \Delta^{\beta}$ and correlation length $\xi \sim \Delta^{-\nu_{\perp}}$, respectively. Our results imply that conserved Manna sandpile, a well studied variant of the CSS, belongs to a distinct universality - {\it not} that of directed percolation (DP), which, without any conservation law as such, does not obey scaling relation (ii).
}
\typeout{polish abstract}
\end{abstract}

\maketitle

\section{Introduction}

Sandpiles \cite{BTW, Kadanoff} were proposed three decades ago as paradigmatic models of ``self-organized criticality'' (SOC) \cite{SOC}, to explain ubiquitous scale-invariant structures in nature. Since then, they continued to capture the imagination of physicists and mathematicians alike \cite{Kardar-Nature1996, Hazra}. Indeed, sandpiles, and the SOC, produced a wealth of results through exact \cite{Dhar-exact, Priezzhev-exact}, numerical \cite{Manna, Vespignani-PRE2000, Rossi-PRL2000, Dickman-PRE2001, Dickman-PRE2002, Lubeck-PRE, Bonachela-PRE2008, Dickman-JSTAT2014, Hexner-PRL2015, Grassberger-PRE2016} and experimental studies \cite{Nagel, Oslo}; for reviews, see  \cite{review}. Yet, by and large, they resisted attempts to construct a unified statistical mechanics framework. In this paper, we discover, in a broad class of conserved-mass sandpiles, a remarkable hydrodynamic structure, which could provide useful insights into large-scale properties of such systems.

Sandpiles are threshold-activated systems of lattice-gases, with sites having non-negative mass or number of particles (or {\it height}). When the number of particles at a site crosses a threshold value, the site becomes active and a fixed number of particles are transferred to its neighbors via toppling. In the original version of sandpile models \cite{BTW}, the system is driven by slowly adding particles, while it relaxes through loss or dissipation of particles at the boundary. On the other hand, in its {\it conserved} ({\it fixed-energy}) version \cite{Vespignani-PRL1998}, total mass remains constant, without any loss or dissipation. However, local bulk-dynamics is the same as in the original sandpiles. Interestingly, upon tuning global density $\rho$, conserved-mass sandpiles undergo an active-absorbing phase transition at a critical density $\rho_c$ \cite{Vespignani-PRL1998, review, APT}. Near criticality, they exhibit scale-invariant structures - reminiscent of that in the original version (dissipative) of sandpiles, without conservation and maintained at criticality through drive and dissipation \cite{review}.

Here we consider only the conserved sandpiles, with stochastic update rules, which we call conserved stochastic sandpiles (CSSs) \cite{Dickman-PRE2001}. Various static and dynamic properties of the CSSs have been studied extensively in the past couple of decades, using simulations \cite{Vespignani-PRE2000, Rossi-PRL2000, Dickman-PRE2001, Dickman-PRE2002, Lubeck-PRE,  Hexner-PRL2015} as well as continuum field theories \cite{Hwa-PRL1989, Munoz-PRL1996, Munoz-PRE2004, Munoz-PRL2005}. However, particle-transport and density-fluctuations, though at the heart of the problem, are far less studied \cite{Carlson, Pradhan-JSTAT2004, Pradhan-PRE2006, Dickman-EPJB2009, Dickman-JSTAT2014, Hexner-PRL2015} and lack general theoretical understanding. Not surprisingly, a long-standing question of universality, fiercely debated over past several decades, is not yet settled \cite{Kadanoff, Biham-PRE1996, Grassberger-JSP1995, Munoz-PRL1996, Mohanty-PRL2002, Munoz-PRE2004, Dickman-PRE2006, Munoz-PRL2005, Munoz-PRL2007, Bonachela-PRE2008, Mohanty-PRL2012, Lee, Dickman-PRE2015, LeDoussal-PRL2015}.
Indeed, it poses a formidable challenge to deal with the issues analytically, precisely because such nonequilibrium many-body systems have nontrivial spatio-temporal correlations; in fact, (quasi-) steady-state probabilities of microscopic configurations are not described by the Boltzmann-Gibbs distributions and are {\it a-priori} not known. In this scenario, one usually resorts to phenomenological field theories, based on symmetries and conservation laws, or to simulations. However, intricacies in simulations make it hard to compare with the theories available and hence to draw a definitive conclusion. Thus, it is highly desirable, and quite pressing at this stage, to understand large-scale properties of sandpiles from the underlying microscopic dynamics itself.

In this paper, we derive hydrodynamics of a broad class of conserved stochastic sandpiles. We demonstrate that these systems, with unbounded state-space, possess a `gradient property' [see Eq. (\ref{GP-DMS})], which we use to uncover a remarkable thermodynamic structure at large scales, with far-reaching consequences: There is an equilibriumlike Einstein relation,
\be
\sigma^2(\rho) = \frac{\chi(\rho)}{D(\rho)},
\label{ER}
\ee 
which connects scaled variance $$ \sigma^2(\rho) = \lim_{v \rightarrow \infty} \frac{ (\langle m^2\rangle - \langle m \rangle^2)}{v}$$ of particle-number, or mass, $m$ in a  subsystem of volume $v$, the bulk-diffusion coefficient $D(\rho)$ and the conductivity $\chi(\rho)$. Here, the density-dependent transport coefficients, bulk diffusion coefficient $D(\rho)$ and conductivity $\chi(\rho)$, are defined from diffusive and drift currents, $$J_{diff} = - D(\rho) \frac{\partial \rho}{\partial x} ~;~ J_{drfit} = \chi(\rho) F,$$ respectively, where $\partial \rho/ \partial x$ is the spatial gradient in density and $F$ is a small force applied in a particular direction. Most crucially, in our theory, the two transport coefficients $D(\rho)$ and $\chi(\rho)$ are related to a macroscopic observable like the activity $a(\rho)$, which is density of active sites in the system. Consequently, probability distributions of subsystem mass is governed by an equilibrium-like chemical potential $\mu(\rho)$, which is also related to the activity density $a(\rho)$.

We demonstrate our results mainly in the context of discrete-mass conserved Manna sandpiles (CMSs) with continuous-time dynamics \cite{Dickman-PRE2001} - a variant of CSSs which has been studied vigorously in the past; we also extend our results to several variants of the CMS as well as to a continuous-mass CSS.
For the CMSs, we have strikingly simple relations between activity $a(\rho)$, bulk-diffusion coefficient $D(\rho) = a'(\rho)$, conductivity $\chi(\rho) = a(\rho)$, and chemical potential $\mu(\rho) = \ln a(\rho)$, leading to an alternative form of the Einstein relation Eq. (\ref{ER}),
\be \sigma^2(\rho) = \frac{a(\rho)}{a'(\rho)}.\ee 
To substantiate our claims, we first directly check the ER in simulations [see Fig. (\ref{ER-fig-DMS})]. Then using the ER, we compute the probabilities of density large-deviations, and the corresponding large deviation functions. In all cases, we find good agreement between theory and simulations.

The above hydrodynamic structure has important consequences on the critical behavior of CSSs. As $\Delta = (\rho - \rho_c) \rightarrow 0^+$, with $\rho_c$ being critical density, we obtain scaling relation (i) where scaled variance $\sigma^2(\rho) \sim (\rho - \rho_c)^{1-\delta}$ with exponent $\delta = 0$; in the special case of a CMS near a critical point, $\sigma^2(\rho) = (\rho - \rho_c)/\beta$ where the proportionality constant is exactly $1/\beta$ (see inset of Fig. \ref{ER-fig-DMS}), with $\beta$ defined from critical behavior of the activity $a(\rho) \sim \Delta^{\beta}$. We obtain another remarkable scaling relation (ii) where dynamical exponent $z = 2 + {(\beta -1)}/{\nu_{\perp}}$ is expressed in terms of two static exponents $\beta$ and $\nu_{\perp}$; exponents $z$ and $\nu_{\perp}$ are defined from critical behaviors of correlation length $\xi \sim \Delta^{-\nu_{\perp}}$ and relaxation time $\tau_r \sim \xi^z$. Indeed, previous estimates of $\beta$, $\nu_{\perp}$ and $z$ in a broad class of the CSS are in reasonably good agreement with scaling relation (ii) [see Table 1 in Sec. \ref{Summary}]. Exponents $\beta$, $\nu_{\perp}$ and $z$ as in Table 1 for directed percolation (DP) \cite{DP, Grassberger-1981}, which have no particle-number conservation as such, violate scaling relation (ii), implying that the conserved Manna sandpiles, and presumably the CSSs in general, belong to a distinct universality, {\it not} that of DP.

\section{Models}

\subsection{Conserved Manna sandpiles (CMS)} 

First we consider conserved Manna sandpiles (CMSs) with continuous-time dynamics. We define the model, for simplicity, on a $d=1$ dimensional ring of $L$ sites (the model can be easily generalized to higher dimensions, other dynamical rules and continuous-mass version of sandpiles as discussed in Secs. \ref{2D}, \ref{PU} and \ref{con-css}). In the CMS, a site $i$ is assigned an {\it unbounded} integer variable $m_i = 0, 1, 2, \dots$, called the number of particles or the mass (such unbounded-mass models are referred also as {\it unrestricted-height} sandpiles).
A site $i$ is {\it active} if $m_i > 1$. The continuous-time dynamics, or equivalently random sequential update (RSU), can be implemented as follows: A site is chosen at random from a list of $N_a$ number of active sites present in the system and is toppled by independently transferring two particles to any of its nearest neighbors, each with {\it equal} probability $1/2$; then, these steps are repeated. For large system sizes, dynamics with RSU (where $\langle N_a \rangle$ sites topple per unit Monte Carlo time) approaches to the continuous-time dynamics \cite{Dickman-PRE2001}. The total number of particles, or mass, $M = \sum_{i=1}^L m_i$ remains {\it conserved}, with density $\rho = M/L$. The activity in the system is measured through active-site density $a(\rho) = \langle N_a \rangle /L$, which depends on density $\rho$, with critical density being $\rho_c \approx 0.95$ \cite{Dickman-PRE2001}.

\subsubsection{Biased conserved Manna sandpiles}

As discussed in the introduction, the conductivity, along with the bulk-diffusion coefficient, plays a crucial role in characterizing fluctuations in CMSs. To calculate the conductivity, we define a generalized (biased) version of a CMS, where there is a constant biasing force $\vec{F}$, coupled to local particle-number and that accordingly modifies the particle-hopping rates in the CMS \cite{Bertini-PRL2001, Jona-Lasinio-RMP2015, Das-PRE2017}. During a toppling in the biased CMS, two particles are transferred, still independently, but each with {\it unequal} probabilities determined according to the transfer-direction of the particle and the magnitude $F$ of the biasing force field $\vec{F} = F \hat{x}$ present along $\hat x$ (similar to stochastic dynamics of a particle of mass $m$ in a gravitational field). So, it is less likely for a particle to go in the direction opposite to the biasing force. The stochastic time-evolution of $m_i(t)$ in the infinitesimal time-interval $dt$ can be written as
\begin{eqnarray}
m_i(t+dt) =
\left\{
\begin{array}{ll} 
m_i(t)-2  & {\rm prob.}~\hat a_i (c_{i,0}^F + c_{i,+}^F+c_{i,-}^F) dt, \cr
m_i(t)+1  & {\rm prob.}~\hat  a_{i-1} c_{i-1,0}^{F} dt, \cr 
m_i(t)+1  & {\rm prob.}~\hat  a_{i+1} c_{i+1,0}^{F} dt, \cr
m_i(t)+2  & {\rm prob.}~\hat  a_{i-1} c_{i-1,+}^{F} dt, \cr
m_i(t)+2  & {\rm prob.}~\hat  a_{i+1} c_{i+1,-}^{F} dt, \cr
m_i(t) & {\rm prob.}~[1-\Sigma dt],
\end{array}
\right.
\label{DM-biased}
\end{eqnarray}
where random variable $\hat a_i = 1$ if a site is active and $\hat a_i=0$ otherwise. The modified (biased) particle-hopping rates  \cite{Bertini-PRL2001, Jona-Lasinio-RMP2015, Das-PRE2017}, 
$$c_{i,\alpha}^F = c^{F=0}_{i,\alpha} \exp \left[ \sum_j \Delta e_{ij}/2 \right],$$ and $\Sigma = [\hat a_i (c_{i,0}^F + c_{i,+}^F + c_{i,-}^F) + \hat a_{i-1} c_{i-1,0}^{F} + \hat a_{i+1} c_{i+1,0}^{F} + \hat a_{i-1} c_{i-1,+}^{F} + \hat a_{i+1} c_{i+1,-}^{F}]$. Here, $\alpha \in \{0,+,-\}$ and the corresponding modified rates $c^F_{i,0}$, $c^F_{i,+}$ and $c^F_{i,-}$ denote  transfer of one particle to the left and one to the right, that of both particles to the right and that of both particles to the left, respectively and $\Delta e_{ij} = \Delta m_{i \rightarrow j} F (j-i)b$ is an `energy cost' \cite{Bertini-PRL2001, Jona-Lasinio-RMP2015, Das-PRE2017} for moving a number $\Delta m_{i \rightarrow j}$ number of particles from site $i$ to $j$ with $b$ being the lattice spacing (for simplicity, we take $b=1$ throughout). The case with $F=0$ corresponds to the unbiased CMS, which is of our interest here. Note that the particle-hopping rates in the unbiased Manna sandpile, which is of our interest in this paper, are actually given by $c_{i,0}^{F=0}=1/2$, $c_{i,+}^{F=0} = c_{i,-}^{F=0} = 1/4$.

\subsubsection{Hydrodynamics}

To calculate conductivity, we expand, as in linear-response theory, the modified rates  in linear order of biasing force $F$ as given below,
\begin{eqnarray}
c^F_{i,0} = c^{F=0}_{i,0} \exp[(F-F)/2] = \frac{1}{2},
\nonumber \\
c^F_{i,+} = c^{F=0}_{i,+} \exp(2 F/2) \simeq \frac{(1+F)}{4},
\nonumber \\
c^F_{i,-} = c^{F=0}_{i,-} \exp(-2 F/2) \simeq \frac{(1-F)}{4}.
\nonumber
\end{eqnarray}
Using the dynamical rules as in Eq. (\ref{DM-biased}), the infinitesimal-time evolution equation for the first moment $\langle m_i \rangle$ of mass at site $i$ can be written as
\begin{eqnarray}
\langle m_i(t+dt) \rangle  
= \langle [m_i(t)-2]\hat a_i \rangle  dt 
+ \langle [m_i(t)+1]\hat a_{i-1} \rangle \frac{dt}{2} \nonumber \\
+ \langle [m_i(t)+1]\hat a_{i+1} \rangle \frac{dt}{2} 
+  (1+F) \langle [m_i(t)+2] \hat a_{i-1} \rangle \frac{dt}{4} \nonumber \\
+ (1-F)\langle [m_i(t)+2] \hat a_{i+1} \rangle \frac{dt}{4} 
+ \langle m_i(t)(1- \Sigma dt) \rangle. \nonumber
\end{eqnarray}
Simplifying the above equation, we obtain the following time-evolution equation for local number-density $\langle m_i(t) \rangle = \rho_i(t)$,
\be
\frac{\partial \rho_i}{\partial t} = (a_{i-1} - 2 a_i + a_{i+1}) + F \frac{a_{i-1} - a_{i+1}}{2}.
\label{GP-DMS}
\ee
where the local activity $\langle \hat a_i \rangle = a_i$. Note that local diffusive current in Eq. (\ref{GP-DMS})  can be expressed as gradient (discrete) of a local observable ($a_i$ here), which we call the `gradient property' \cite{Jona-Lasinio-RMP2015}. As discussed below, the gradient property helps one to immediately identify the bulk-diffusion coefficient and conductivity in the CMS.

Now, from a simple physical consideration in a large system on a macroscopic scale, where density and activity fields are slowly varying functions of space and time, it would be quite reasonable to assume that the local activity is {\it not independent}, but rather {\it ``slave''} to the local density. This is because, on the coarse-grained level where a hydrodynamic theory is valid, the relaxation time-scales, within a subsystem, for a conserved density field (a {\it ``slow''} variable) and a nonconserved activity field (a {\it ``fast''} variable) are expected to be separated and, therefore, local activity should take the value corresponding to the instantaneous local density.
In other words, we assume here a property that there exists a local steady-state, where the average of any local observable $g(m_i)$ could be replaced by its steady-state average $\langle g(m_i) \rangle = \langle g(m_i) \rangle^{st}_{\rho_i}$ corresponding to the local density $\rho_i$  \cite{ Eyink, Bertini-PRL2001, Jona-Lasinio-RMP2015, Das-PRE2017}. Here, in our case, writing $g(m_i) \equiv \hat a_i = (1-\delta_{m_i,0} - \delta_{m_i,1})$, we could express local activity as $a_i = \langle \hat a_i \rangle^{st}_{\rho_i} \equiv a[\rho_i(t)]$, which is now a function of only local density $\rho_i(t)$. It may be noted that this particular property is somewhat analogous to that of local equilibrium, where the average of any local observable on a macroscopic scale is equal to an equilibrium average, which is calculated w.r.t. the Boltzmann-Gibbs distribution corresponding to the local value of the density. However, the average of the local observable, i.e., the local activity, should be calculated here in a nonequilibrium (quasi-) steady state, {\it not} in an equilibrium one; for a nice exposition of the concept of local steady state, we refer the reader to Refs. \cite{Eyink, Bertini-PRL2001, Jona-Lasinio-RMP2015}. Consequently, after taking the continuum limit where one rescales space $i \rightarrow x=i/L$ (therefore, also rescaling lattice spacing $b \rightarrow 1/L$) and time $t \rightarrow t/L^2$, Eq. (\ref{GP-DMS}) leads to the desired hydrodynamic evolution of the density field $\rho(x,t)$ at position $x$ and time $t$ as given below,
\be 
\frac{\partial \rho(x, t)}{\partial t} = \frac{\partial^2 a(\rho)}{\partial x^2} - F \frac{\partial a(\rho)}{\partial x} \equiv - \frac{\partial J}{\partial x}.
\label{Hyd-biased}
\ee
From the above hydrodynamic evolution Eq. (\ref{Hyd-biased}), which is nothing but the continuity equation for locally conserved mass, the local current $J(\rho(x)) = J_{diff} + J_{drift}$ can be immediately constructed through standard prescription: The diffusion current, $$J_{diff} \equiv - D(\rho) \frac{\partial \rho}{\partial x},$$ and the drift current, $$J_{drift} \equiv \chi(\rho) F,$$ with the density-dependent bulk-diffusion coefficient $D(\rho)$ and conductivity $\chi(\rho)$ can be identified as $D(\rho) = {da}/{d\rho} \equiv a'(\rho)$ and $\chi(\rho) = a(\rho)$, respectively.

The hydrodynamic structure as derived in Eq. (\ref{Hyd-biased}) is the first important result of this paper, and constitutes the basis of the whole analysis here. Indeed, there are certain similarities between hydrodynamic equation (\ref{Hyd-biased}) and the previously obtained coarse-grained field theories \cite{Munoz-PRL1996, Vespignani-PRL1998, Munoz-PRL2005}. However, there are important differences too: Eq. (\ref{Hyd-biased}) here involves only a single field variable, i.e., conserved density field $\rho(x,t)$; the activity $a[\rho(x,t)]$, as explained above, is not treated here as an independent field variable, rather it evolves through its (nonlinear) dependence on density field.

A mathematically rigorous proof of the existence of hydrodynamic limit for interacting-particle systems is a fundamental problem and, quite remarkably, have been carried out in the past for several models \cite{Eyink, Carlson2, Landim}. However, rigorously establishing hydrodynamic equation (\ref{Hyd-biased}) for nontrivial models like sandpiles, for which microscopic structure is not exactly known, is technically challenging and presently beyond the scope of this work. Importantly though, the hydrodynamics as in Eq. (\ref{Hyd-biased}), and the local steady-state property used to derive it, can be readily tested by verifying its remarkable consequences, which are discussed in the following sections.

\subsubsection{Macroscopic fluctuation theory (MFT)}
\label{Sec-MFT}

Following the prescription of macroscopic fluctuation theory (MFT) \cite{Bertini-PRL2001, Jona-Lasinio-RMP2015}, we now use the above hydrodynamics Eq. (\ref{Hyd-biased}) to characterize fluctuations, on a coarse-grained level, in the {\it unbiased} system with $F=0$ (the actual CMS), solely in terms of the two density-dependent transport coefficients - the bulk-diffusion coefficient $D(\rho)$ and conductivity $\chi(\rho)$. To elaborate on this point, we consider a system, which is divided into $\nu=L/l$ large subsystems of size $l \ll L$. Then the joint probability distribution ${\cal P}[\hat \rho_1, \hat \rho_2, \dots, \hat \rho_{\nu}]$ of the subsystem number-densities $\{\hat \rho_{\alpha} = M_{\alpha}/l \}$, with $M_{\alpha}$ being the particle-number in the $\alpha$th subsystem and $\alpha \in \{1,2, \dots, \nu\}$, can be written as a product of subsystem weight-factors \cite{Chatterjee-PRL2014, Chatterjee-PRE2016},
\be 
{\cal P}[\{\rho_{\alpha}\}] \simeq \prod_{\alpha} \exp[- \{f(\hat \rho_{\alpha} -f(\rho) - \mu(\rho) (\hat \rho_{\alpha} -\rho) \}], 
\label{additivity}
\ee 
with $\rho=M/L$ being the global density. In a suitable coarse-grained limit, the joint distribution can also be written as ${\cal P} \simeq \exp\{-{\cal H}[\hat \rho(x)]\}$ where ${\cal H}[\hat \rho(x)] = \int dx [f(\hat \rho(x)) - f(\rho) - \mu(\rho) (\hat \rho(x)-\rho)]$ with $f(\rho)$ and $\mu(\rho)=df/d\rho$ being equilibrium-like free-energy density and chemical potential, respectively. According to the MFT, free-energy density $f(\rho)$ can be determined by solving a Hamilton-Jacobi equation \cite{Bertini-PRL2001, Jona-Lasinio-RMP2015},
\be 
\int dx \frac{\partial}{\partial x} \left(  \frac{\delta {\cal H}}{\delta \hat \rho} \right) \chi(\hat \rho) \frac{\partial}{\partial x} \left( \frac{\delta {\cal H}}{\delta \hat \rho} \right) - \int dx \frac{\delta {\cal H}}{\delta \hat \rho} \frac{\partial J_{diff}}{\partial x} = 0,
\label{MFT}
\ee
implying $f''(\rho) = (d\mu/d\rho)= D(\rho)/\chi(\rho)$. Now, as the product form of joint subsystem mass distribution ${\cal P}[\{\hat \rho_{\alpha} = M_{\alpha}/l\}]$ in Eq. (\ref{additivity}) implies a fluctuation-response relation (FR) \cite{Chatterjee-PRL2014}, involving an equilibrium-like compressibility $d\rho/d\mu$ and mass fluctuation $\sigma^2(\rho) = \lim_{l \rightarrow \infty} (\langle M_{\alpha}^2 \rangle - \langle M_{\alpha} \rangle^2)/l$,
\be \frac{d\rho}{d\mu} = \sigma^2(\rho),
\label{FR}
\ee 
we arrive at the Einstein relation (ER) as in Eq. (\ref{ER}). Remarkably, in the unbiased CMS with the biasing force $F=0$, we obtain, using the exact relations $D(\rho) = a'(\rho)$ and $\chi(\rho) = a(\rho)$, an alternative form of the ER, directly connecting mass fluctuation to activity, 
\be 
\sigma^2(\rho) = \frac{a(\rho)}{a'(\rho)}.
\label{ER-DMS}
\ee 
The above relation in Eq. (\ref{ER-DMS}) is the second important result of this paper (for other variants of the CSS, see Secs. \ref{2D}, \ref{PU} and \ref{con-css}). By integrating  $d\mu/d\rho = a'(\rho)/a(\rho)$, obtained from Eqs. (\ref{FR}) and (\ref{ER-DMS}), one immediately has the chemical potential as given below, \be \mu(\rho) = \ln a(\rho) + c, \label{mu_cms} \ee with $c$ being an arbitrary constant of integration.

\begin{figure}[h]
\includegraphics[scale=0.72]{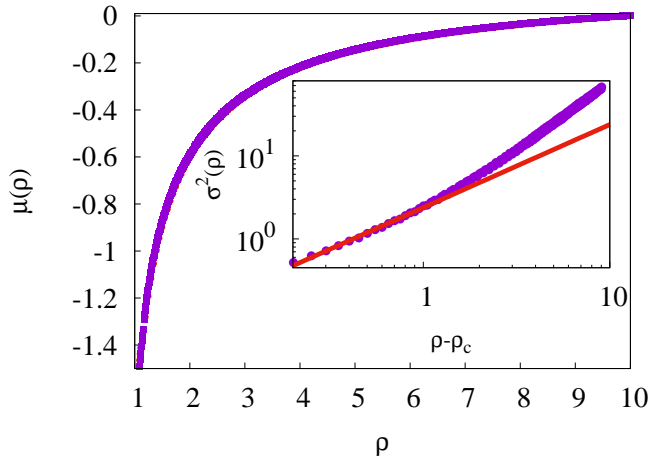}
\caption{Chemical potentials $\mu(\rho) = \int_{\rho_{_0}}^{\rho} a'(\rho)/a(\rho) d\rho$ [red circles; integrating the inverse of the rhs of Eq. (\ref{ER-DMS})] and $\mu(\rho) = \int_{\rho_{_0}}^{\rho} 1/\sigma^2(\rho) d \rho$ [magenta squares; integrating the inverse of the lhs of Eq. (\ref{ER-DMS})] are plotted as a function of density $\rho$. {\it Inset:} The scaled variance $\sigma^2(\rho)$ {\it vs.} $(\rho-\rho_c)$, is plotted (magenta circles) where red line [theory, scaling relation (i)] represents $\sigma^2=(\rho-\rho_c)/\beta$, with $\rho_c \approx 0.95$ and $\beta \approx 0.42$ \cite{Dickman-PRE2001}. }
\label{ER-fig-DMS}
\end{figure}

In the following, we check the integrated form of the ER Eq. (\ref{ER-DMS}) [or, equivalently, Eq. (\ref{ER})]. We first calculate from simulations both the scaled variance $\sigma^2(\rho)$ and the activity $a(\rho)$ as a function of density $\rho$. Then we calculate chemical potential in two ways: $\mu(\rho) = \int_{\rho_0}^{\rho} 1/\sigma^2 d\rho$ by integrating the inverse of the lhs of Eq, (\ref{ER-DMS}) and $\mu(\rho) = \int_{\rho_0}^{\rho} a'(\rho)/a(\rho) d\rho = [\ln a(\rho) - \ln a(\rho_0)]$ by integrating the inverse of the rhs of Eq. (\ref{ER-DMS}). In Fig. \ref{ER-fig-DMS}, both the chemical potentials are plotted  as a function of density $\rho$ and, and as seen in the plot, they are in excellent agreement with each other, thus verifying the ER for the CMS.

\subsubsection{Scaling relations}

The above hydrodynamic structure has two important consequences on the near-critical behavior of the CMSs. 
\\
\\
(i) Since the chemical potential $\mu(\rho) = \ln a(\rho)+c$ as in Eq. (\ref{mu_cms}) and the activity $a(\rho) \sim \Delta^{\beta}$ as $\Delta = (\rho - \rho_c) \rightarrow 0^+$, we immediately obtain scaling relation (i) by using Eq. (\ref{FR}) [or alternatively, by using Eq. (\ref{ER-DMS})]: The scaled variance $$ \sigma^2(\rho) \simeq {\rm Const}. \Delta^{1-\delta}$$ of subsystem-mass near criticality is proportional to $(\rho-\rho_c)$, with exponent \be \delta=0 \ee and the proportionality constant is exactly $1/\beta$, $\sigma^2(\rho) = (\rho - \rho_c)/\beta.$
In the inset of Fig. \ref{ER-fig-DMS}, we plot scaled variance $\sigma^2(\rho)$ of subsystem mass as a function of $(\rho-\rho_c)$, which is in quite good agreement with simulations. \\
\\
(ii) We obtain the second scaling relation as follows. First we note that, in the unbiased CMS, Eq. (\ref{Hyd-biased}) with vanishing biasing force ($F=0$) leads to the hydrodynamic evolution equation $$\frac{\partial \rho(x,t)}{\partial t} = \frac{\partial}{\partial x} \left[ D(\rho) \frac{\partial \rho}{\partial x} \right]. $$ Now, a simple dimensional analysis would imply the relaxation time $\tau_r \sim \xi^2/D \sim \xi^{2 + (\beta-1)/\nu_{\perp}}$ where the spatial correlation length $\xi \sim \Delta^{-\nu_{\perp}}$ and the bulk-diffusion coefficient $D(\rho) = a'(\rho) \sim \Delta^{\beta-1} \sim \xi^{(1-\beta)/\nu_{\perp}}$, all of them diverging at criticality. Defining the dynamic exponent $z$ as $\tau_r \sim \xi^z$, we obtain scaling relation (ii), \be z = 2 + \frac{(\beta-1)}{\nu_{\perp}},\ee where dynamic exponent $z$ is expressed in terms of static exponents $\beta$ and $\nu_{\perp}$, reminiscent of similar relations in equilibrium critical phenomena \cite{Halperin-Hohenberg}.

\subsubsection{Density large deviations} 

We can numerically compute, by integrating the FR Eq. (\ref{FR}) to obtain $\mu(\rho)$ and $f(\rho)$ \cite{Subhadip}, the probability  of large deviation, \be P_v(m) \sim e^{- v h(\hat \rho)}, \ee where $\hat \rho = m/v$ is coarse-grained density defined over a subsystem of volume $v=l^d$ in $d$ dimensions and the large deviation function (LDF) can be written as $h(\hat \rho) = f(\hat \rho) -\mu(\rho) \hat \rho$, with $\rho$ being the global density. However, the LDF has sub-leading corrections, which can also be obtained here by taking the asymptotic form of $\sigma^2(\rho) \simeq \rho^2/\eta$, which is actually the case in the limit of large density $\rho \gg \rho_c$, with $\eta$ a model-dependent proportionality constant. This particular asymptotic form of mass fluctuation is quite expected as the CMS, being defined in a unbounded state-space, behaves somewhat like a `Bose gas', with repulsive interactions. This form of $\sigma^2(\rho)$ then implies Legendre-Fenchel transform $\lambda_v(\kappa) = {\rm \bf inf}_{\rho}[f(\rho)-\kappa \rho]$ of free energy density $f(\rho)$, to have an expression $\lambda_v(\kappa) \simeq {\rm const.} - v \eta \ln \kappa$, which is valid for small $\kappa$ or, equivalently, for large subsystem mass $m$. Consequently, to leading order of $m$ and $v$, we obtain $f(\hat \rho) \simeq v [\eta \ln (\hat \rho) - \ln m/v]$, implying a sub-leading correction to the leading-order term $\eta \ln (\hat \rho)$. In other words, we have $P_v(m) \simeq {\rm const.} \exp[-v h(\hat \rho)]/m$ with a $1/m$ correction at large $m$, which means that the large-mass behavior of $P_v(m)$, even at lower densities, is essentially determined by the large-density behavior of mass fluctuations $\sigma^2(\rho)$.

\begin{figure}[h]
\includegraphics[scale=0.7]{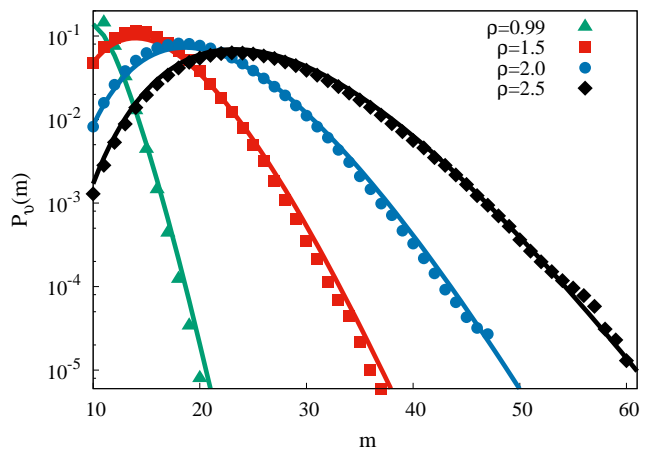}
\caption{Probability $P_v(m) \simeq {\rm const.} \exp[-v h(\hat \rho)]/m$ of large deviation in subsystem mass $m$, with the LDF $h(\hat \rho) = f(\hat \rho) -\mu(\rho) \hat \rho$ computed exactly by numerically integrating Eq. (\ref{FR}), is plotted as a function of $m$ for densities $\rho = 0.99$ (green triangles), $1.5$ (red squares), $2.0$ (blue circles) and $2.5$ (black diamonds). Points - simulations, lines - theory.}
\label{LDF-fig-DMS}
\end{figure}

In Fig. \ref{LDF-fig-DMS}, for CMS in $d=1$ dimension, we plot the subsystem mass-distributions $P_v(m)$ as a function of subsystem mass $m$ for various densities $\rho = 0.99, 1.5, 2.0$ and $2.5$. Simulations (points) are in reasonably good agreement, especially at larger densities, with theoretically obtained subsystem mass distributions (lines); the small deviations observed are presumably due to the finite size effects, which come into play near criticality.

\subsubsection{Comparison with mean-field theory}

To stress that the above hydrodynamic theory indeed captures nontrivial correlations present in the systems, here we perform a mean-field analysis, which ignores spatial correlations, and then we compare the mean-field results with that obtained using our hydrodynamic theory.

\begin{figure}[h]
\includegraphics[scale=0.7]{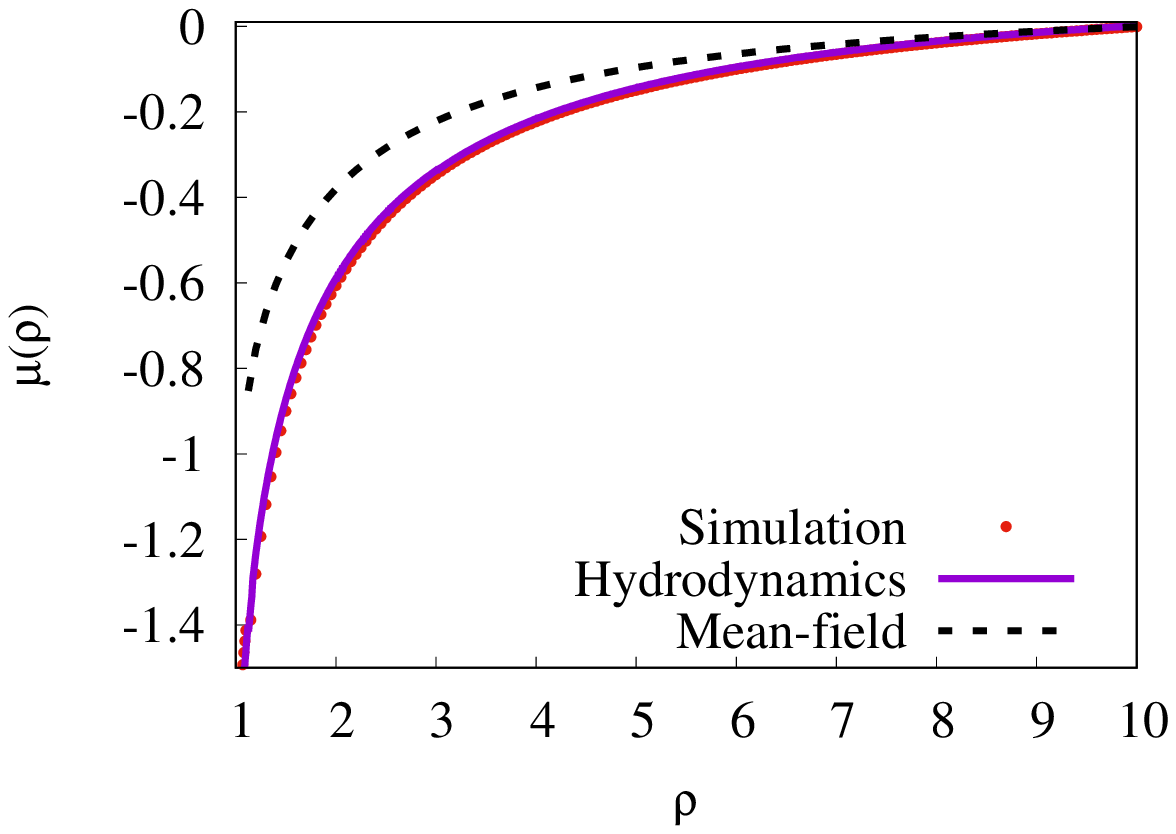}
\includegraphics[scale=0.68]{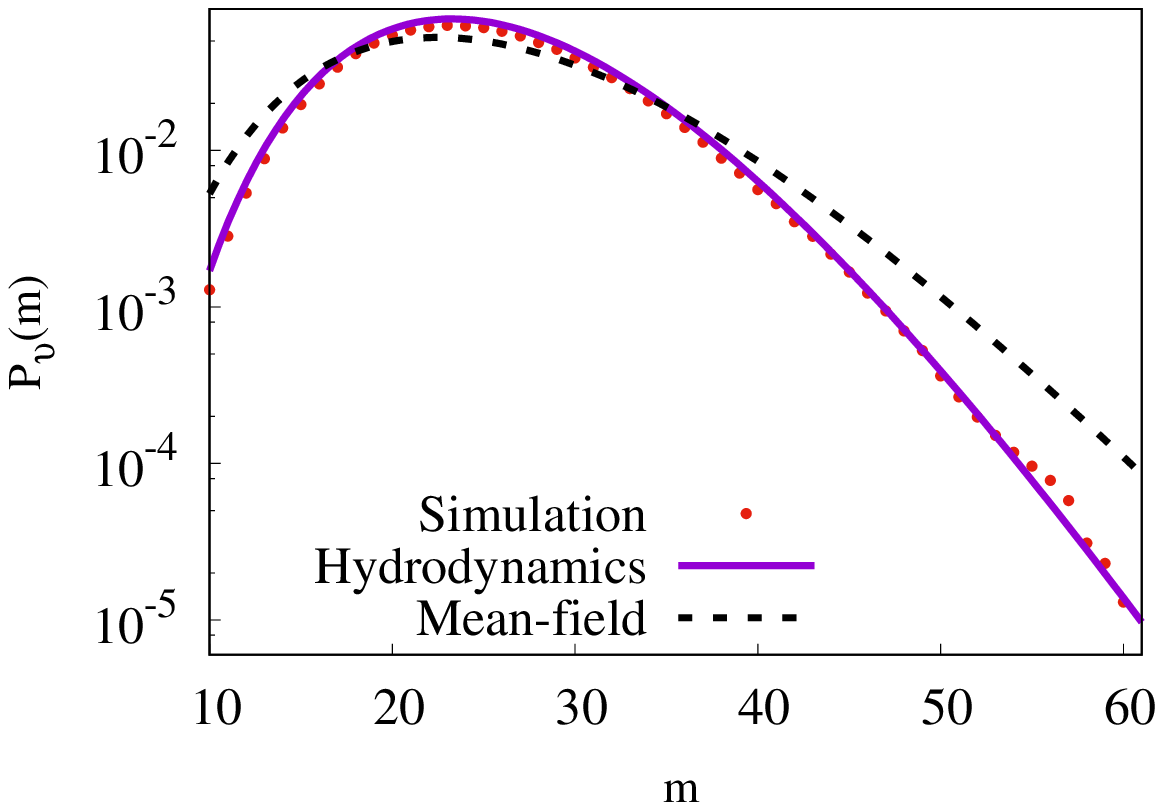}
\caption{ Top panel: Chemical potential $\mu(\rho)$ is obtained through $\mu(\rho) = \int_{\rho_0}^{\rho} 1/\sigma^2 d\rho$ (simulations), $\mu(\rho) = \int_{\rho_0}^{\rho} D(\rho)/\chi(\rho) d\rho$ and $\mu(\rho) = \int_{\rho_0}^{\rho} 1/\sigma^2 d\rho$ [mean-field $\sigma^2$, somewhat improved though, from Eq. \ref{mftsig} by estimating $p(\rho)$ from simulations]. Bottom panel: We compare subsystem mass distributions $P_v(m)$ obtained from simulations and hydrodynamic theory to that obtained from the mean-field theory (here, density value $\rho=2.5$ is used). Simulations - red circles, mean field theory - black dotted lines, hydrodynamic theory - violet lines. Clearly, hydrodynamic theory captures the simulation results remarkably well and significantly better than the mean-field theory. }
\label{hyd-MF}
\end{figure}

The time evolution equation for the $n$-th moment $\langle m_i^n \rangle$ of mass at site $i$ can be written as
\begin{eqnarray}
\langle m_i^n(t+dt) \rangle     
= \langle [m_i(t) -2]^n \hat a_i \rangle  dt \nonumber + \langle [m_i(t)+1]^n \hat a_{i-1} \rangle \frac{dt}{2} \nonumber \\
+ \langle [m_i(t)+1]^n \hat a_{i+1} \rangle \frac{dt}{2} +  \langle [m_i(t)+2]^n \hat a_{i-1} \rangle \frac{dt}{4} \nonumber \\
+ \langle [m_i(t)+2]^n \hat a_{i-1} \rangle \frac{dt}{4} \nonumber \\
+ \langle m_i^n(t) [1-(\hat a_i + \frac{3}{4} \hat a_{i-1} + \frac{3}{4} \hat a_{i+1})dt)] \rangle. ~~~~~\label{nthmoment}
\end{eqnarray}
In the steady state, we must have $\langle m_i^n(t+dt) \rangle = \langle m_i^n(t) \rangle$, i.e. $d \langle m_i^n(t) \rangle/dt=0$. Note that there are indeed nontrivial correlations between local mass $m_i$ and activity $\hat a_j$ (nearest neighbors $j=i\pm1$) as they are coupled in Eq. \ref{nthmoment}. Now putting $n=2$ in Eq. \ref{nthmoment} and using the steady-state condition $\langle m_i^2(t+dt) \rangle = \langle m_i^2(t) \rangle$, we get
\begin{eqnarray}
\langle m_i^2\rangle = [\langle m_i^2 \hat a_i \rangle - 4 \langle m_i \hat a_i \rangle + 4 \langle \hat a_i \rangle \nonumber] dt +
 [\langle m_i^2 \hat a_j \rangle + 2 \langle m_i \hat a_j \rangle \nonumber \\
 +\langle \hat a_j \rangle] dt +
 [\langle m_i^2 \hat a_j \rangle + 4 \langle m_i \hat a_j \rangle + 4 \langle \hat a_j \rangle] dt/2 \nonumber \\
 + \langle m_i^2 \rangle - \langle m_i^2 (\hat a_i + \frac{3}{4} \hat a_{i-1} + \frac{3}{4} \hat a_{i+1}) \rangle dt, ~~
 \nonumber
\end{eqnarray}
where nearest neighbors $j=i\pm1$ and the second moment $\langle m_i^2 \rangle$ cancels out. Then defining activity $\langle \hat a_i \rangle = \langle (1- \delta_{m_i,0} - \delta_{m_i,1}) \rangle \equiv a(\rho)$ as the probability that a site is atleast doubly occupied or active, and using the mean-field approximation that two-point correlations vanish for neighboring sites $i \ne j$ (i.e., $\langle m_i \hat a_j \rangle = \langle m_i \rangle \langle \hat a_j \rangle$ and $\langle m_i^2 \hat a_j \rangle = \langle m_i^2 \rangle \langle \hat a_j \rangle$) and $\langle m_i \rangle =\rho, \langle m_i \delta_{m_i,1} \rangle = p(\rho) \equiv {\rm Prob.}(m_i=1)$, we get
\begin{equation}
a(\rho) = \frac{4(\rho-p)}{(7+4\rho)}. \label{s2}
\end{equation}
Following the above procedure for $n=3$, we get, from Eq. \ref{nthmoment}, 
\begin{eqnarray}
\langle m_i^3 \rangle = [\langle m_i^3 \hat a_i \rangle - 6 \langle m_i^2 \hat a_i \rangle + 12 \langle m_i \hat a_i \rangle -8 \langle \hat a_i \rangle] dt 
\nonumber \\
+[\langle m_i^3 \hat a_j \rangle + 3 \langle m_i^2 \hat a_j \rangle + 3 \langle m_i \hat a_j \rangle + \langle \hat a_j \rangle] dt 
\nonumber \\
+ [\langle m_i^3 \hat a_j \rangle + 6 \langle m_i^2 \hat a_j \rangle + 12 \langle m_i \hat a_j \rangle + 8 \langle \hat a_j \rangle] \frac{dt}{2} 
\nonumber \\ 
+ \langle m_i^3 \rangle - \langle m_i^3 (\hat a_i + \frac{3}{4} \hat a_{i-1} + \frac{3}{4} \hat a_{i+1} ) \rangle dt,
\nonumber
\end{eqnarray}
leading to the mean-field expression for the second moment ($m_i^3$ cancels out on the level $n=3$),
$$
\langle m_i^2 \rangle=\frac{(14 \rho^2 -10 p \rho + 12 \rho- 5 p)}{(7 + 4 p)},
$$
and, accordingly, the variance $\sigma^2=\langle m_i^2 \rangle -\rho^2$ as given below
\begin{eqnarray}
\sigma^2(\rho) &=& \frac{(7-4p)\rho^2+(12-10 p)\rho - 5 p}{(7 + 4 p)}.
\label{mftsig}
\end{eqnarray}
Note that, even on the mean-field level, it is still not easy to solve the infinite hierarchy involving various moments as determining $p(\rho)$ as a function of $\rho$ requires solution of the full hierarchy (see Eq. \ref{nthmoment}), which will be addressed elsewhere. However, it is easy to check the large-density limit of the fluctuation, which is given by $\sigma^2(\rho) \simeq \rho^2$ where we use that ${\rm Prob.}(m_i=1) = p(\rho)$ should vanish at large $\rho$. For conserved Manna sandpiles (and CSSs in general), two-point spatial correlations, which have been ignored in the above mean-field analysis, are actually nonzero at all densities (even at large density) and therefore cannot capture the actual fluctuations obtained from simulations. On the other hand, our hydrodynamic theory nicely captures these correlations and consequently captures the actual fluctuations in the system remarkably well. The deviation from the mean-field theory and the agreement between hydrodynamic theory and simulations are quite evident in Fig. \ref{hyd-MF} where simulation, hydrodynamics and mean-field results for scaled variance $\sigma^2(\rho)$ as a function of density $\rho$ and subsystem mass distribution $P_v(m)$, at a particular density $\rho=2.5$, as a function of subsystem mass $m$ have been compared.


\subsubsection{Higher dimensions}
\label{2D}

Here we generalize the biased version of a conserved Manna sandpile with continuous-time dynamics (random sequential updates) on a two dimensional (2D) lattice. In the presence of a biasing force $\vec{F} = F \hat{x}$ in a particular direction (anti-clockwise, say), the modified particle-hopping \cite{Bertini-PRL2001, Jona-Lasinio-RMP2015, Das-PRE2017} rates can be written as $$c_{i, j, \alpha, \alpha'}^F = c^{F=0}_{i,j,\alpha,\alpha'} \exp \left[ \sum_{(i',j')} F \Delta m_{(i,j) \rightarrow (i',j')}/2 \right],$$
where $c_{i, j, \alpha, \alpha'}^F$ is the rate for the toppling event at an active site $(i,j)$ where one particle goes to the $\alpha$th direction and the other to the $\alpha'$th direction, with $\alpha, \alpha' \in \{+\hat{x}, -\hat{x}, +\hat{y}, -\hat{y} \}$. The rates can be explicitly written in linear order of biasing force $F$ as given below,
\begin{eqnarray}
\nonumber 
C^{F}_{i,j,+x,+x} &=& C^{F=0}_{i,j,+x,+x} \exp(2F/2) \simeq \frac{(1+F)}{16},
\\ \nonumber
C^{F}_{i,j,-x,-x} &=& C^{F=0}_{i,j,-x,-x} \exp(-2F/2) \simeq \frac{(1-F)}{16},
\\ \nonumber
C^{F}_{i,j,+y,+y} &=& C^{F=0}_{i,j,+y,+y} = \frac{1}{16},
\\ \nonumber
C^{F}_{i,j,-y,-y} &=& C^{F=0}_{i,j,-y,-y} = \frac{1}{16},
\\ \nonumber
C^{F}_{i,j,+x,-x} &=& C^{F=0}_{i,j,+x,-x} \exp[(F-F)/2] = \frac{1}{8},
\\ \nonumber
C^{F}_{i,j,+x,\pm y} &=& C^{F=0}_{i,j,+x,\pm y} \exp(F/2) \simeq \frac{(1+F/2)}{8},
\\ \nonumber
C^{F}_{i,j,-x,\pm y} &=& C^{F=0}_{i,j,-x,\pm y} \exp(-F/2) \simeq \frac{(1-F/2)}{8},
\\ \nonumber
C^{F}_{i,j,+y,-y} &=& C^{F=0}_{i,j,+y,-y} = \frac{1}{8}.
\end{eqnarray}

\begin{figure}
\includegraphics[scale=0.7]{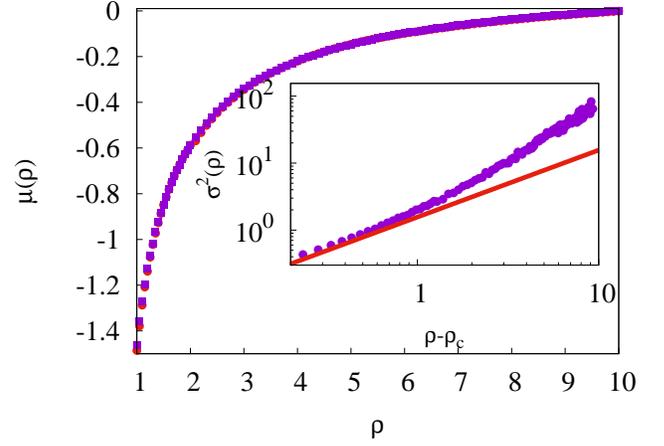}
\caption{Verification of Einstein relation (ER) in the conserved Manna sandpile with continuous-time (random sequential) update in two dimensions (2D): Chemical potentials $\mu(\rho) = \int_{\rho_0}^{\rho} D(\rho)/\chi(\rho) d\rho$, with $D(\rho) = (1/2) d a(\rho)/d\rho \equiv a'(\rho)/2$ and $\chi(\rho)=a(\rho)/2$, and $\mu(\rho) = \int_{\rho_0}^{\rho} 1/\sigma^2(\rho) d\rho$ are plotted as function of density $\rho$ [obtained by integrating the inverse of the rhs and the lhs of Eq. (1) in the main text]. {\it Inset:} Scaled variance $\sigma^2(\rho)$ {\it vs.} $(\rho-\rho_c)$, is plotted (magenta circles) where red line [theory, scaling relation (i)] represents $\sigma^2=(\rho-\rho_c)/\beta$, with estimated critical density $\rho_c \approx 0.72$ and exponent $\beta \approx 0.64$ \cite{Vespignani-PRE2000}. }
\label{ER-2D}
\end{figure}

We define a set of random variable $\hat a_{i,j} = 1$ if a site is active and $\hat a_{i,j} = 0$ otherwise. Then, the time-evolution of mass $m_{i,j}(t)$ at site $(i,j)$ at time $t$ can be written in the infinitesimal time-interval $dt$,
\begin{eqnarray}
m_{i,j}(t+dt) =
\left\{
\begin{array}{ll} 
m_{i,j}(t)-2  & {\rm prob.}~\hat a_{i,j} dt, \cr
m_{i,j}(t)+1  & {\rm prob.}~\hat a_{(i-1),j} \frac{(3+F)}{8} dt, \cr 
m_{i,j}(t)+1  & {\rm prob.}~\hat a_{(i+1),j} \frac{(3-F)}{8} dt, \cr 
m_{i,j}(t)+1  & {\rm prob.}~\hat a_{i,(j-1)} \frac{3}{8} dt, \cr 
m_{i,j}(t)+1  & {\rm prob.}~\hat a_{i,(j+1)} \frac{3}{8} dt, \cr 
m_{i,j}(t)+2  & {\rm prob.}~\hat a_{(i-1),j} \frac{(1+F)}{16} dt, \cr 
m_{i,j}(t)+2  & {\rm prob.}~\hat a_{(i+1),j} \frac{(1-F)}{16} dt, \cr 
m_{i,j}(t)+2  & {\rm prob.}~\hat a_{i,(j-1)} \frac{1}{16} dt, \cr 
m_{i,j}(t)+2  & {\rm prob.}~\hat a_{i,(j+1)} \frac{1}{16} dt, \cr 
m_{i,j}(t)    & {\rm prob.}~[1-\Sigma dt]
\end{array}
\right.
\end{eqnarray}
where 
\begin{eqnarray}
\Sigma = \hat a_{i,j}  + \hat a_{(i-1),j} \frac{(3+F)}{8}  + \hat a_{(i+1),j} \frac{(3-F)}{8}  + \hat a_{i,(j-1)} \frac{3}{8} 
\nonumber \\
+ \hat a_{i,(j+1)} \frac{3}{8} + \hat a_{(i-1),j} \frac{(1+F)}{16} + \hat a_{(i+1),j} \frac{(1-F)}{16}
\nonumber \\ + \hat a_{i,(j-1)} \frac{1}{16} 
+ \hat a_{i,(j+1)} \frac{1}{16}.
\nonumber
\end{eqnarray} 
The local density variable $\rho_{i,j}(t) = \langle m_{i,j} (t) \rangle$ at site $(i,j)$ and time $t$ evolves through the following equation, 
\begin{eqnarray}
\frac{d \rho_{i,j}}{dt} &=& \frac{1}{2} \left[ a_{(i+1),j} - 2 a_{i,j} + a_{(i-1),j} \right] 
\nonumber \\ 
&+& \frac{1}{2} \left[ a_{i,(j+1)} - 2 a_{i,j} 
+ a_{i,(j-1)} \right] 
\nonumber \\
&-& \frac{F}{4}  \left[ a_{(i+1),j} -  a_{(i-1),j} \right],
\end{eqnarray}
which clearly has the gradient property. Now, in the continuum limit by rescaling space $\{i,j\} \rightarrow \{x=i/L, y=j/L\}$ and time $t \rightarrow t/L^2$ and using the property of local steady state discussed previously, the above equation leads to the desired hydrodynamic evolution equation for density field $\rho(\vec{r},t)$ at position $\vec{r}=\{x,y\}$ and time $t$,
\be 
\frac{\partial \rho(\vec{r}, t)}{\partial t} = \frac{1}{2} \nabla^2 a(\rho) -\frac{1}{2} F \frac{\partial a(\rho)}{\partial x} \equiv - \nabla.\vec{J}(\rho(\vec{r})).
\ee
In the above hydrodynamic equation, the local current $\vec{J} = \vec{J}_{diff} + \vec{J}_{drift}$ has two parts: diffusive current $\vec{J}_{diff} = - (1/2) \nabla a(\rho) \equiv - D(\rho) \nabla \rho$ and drift current $\vec{J}_{drift} = \chi(\rho) \vec{F}$
where bulk-diffusion coefficient $D(\rho) = a'(\rho)/2$ and conductivity $\chi = a(\rho)/2$; the two transport coefficients are density-dependent in general. Now, following macroscopic fluctuation theory \cite{Bertini-PRL2001, Jona-Lasinio-RMP2015} as discussed in Sec. \ref{Sec-MFT} in the case of one dimensional CMS, we recover the Einstein relation $\sigma^2(\rho) = {\chi(\rho)}/{D(\rho)} = {a(\rho)}/{a'(\rho)}$ also in two dimensional CMS, or alternatively,
\begin{equation}
\sigma^2(\rho)  = \left[ \frac{d(\ln a)}{d \rho} \right]^{-1}.
\label{ER-FR-2D}
\end{equation}
By integrating the above equation, we immediately obtain an equilibrium-like chemical potential, 
\be 
\mu(\rho)  = \ln a(\rho) - \ln a(\rho_{_0}),
\ee
which would now govern the coarse-grained density fluctuation in a subsystem. Note that mass-fluctuation $\sigma^2(\rho)$ and chemical potential $\mu(\rho)$ in the one and two dimensional CMS have the same dependence on the activity $a(\rho)$. In Fig. (\ref{ER-2D}), we have plotted the chemical potentials, $\mu(\rho) = \int_{\rho_0}^{\rho} 1/\sigma^2 d\rho$ by integrating inverse of lhs of Eq. (\ref{ER-FR-2D}) and $\mu(\rho) = [\ln a(\rho) - \ln a(\rho_0)]$ by integrating the inverse of the rhs of Eq. (\ref{ER-FR-2D}), as a function of density $\rho$. As seen in the plot, the two chemical potentials are in excellent agreement with each other, thus verifying the ER for the CMS in two dimensions.

\subsubsection{Parallel updates (PU)}
\label{PU}

\begin{figure}
\includegraphics[scale=0.7]{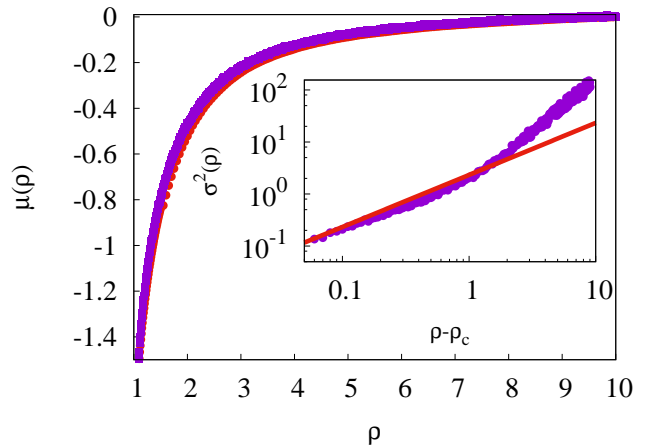}
\caption{Verification of Einstein relation (ER) in the conserved Manna sandpile with parallel update (PU) in one dimension: Chemical potentials $\mu(\rho) = \int_{\rho_0}^{\rho} D(\rho)/\chi(\rho) d\rho$, with $D(\rho) = d a(\rho)/d\rho \equiv a'(\rho)$ and $\chi(\rho)=a(\rho)$, and $\mu(\rho) = \int_{\rho_0}^{\rho} 1/\sigma^2(\rho) d\rho$ are plotted as function of density $\rho$ [obtained by integrating the inverse of the rhs and the lhs of Eq. (1) in the main text]. {\it Inset:} Scaled variance $\sigma^2(\rho)$ {\it vs.} $(\rho-\rho_c)$, is plotted (magenta circles) where red line [theory, scaling relation (i)] represents $\sigma^2=(\rho-\rho_c)/\beta$, with estimated critical density $\rho_c \approx 0.92$ and exponent $\beta \approx 0.43$. }
\label{ER-fig-DMS-PU}
\end{figure}

In this section, we consider one dimensional conserved Manna sandpile with parallel update rules, where all active sites are updated simultaneously at each discrete time step; generalization to higher dimension is straightforward. Let us define two random variables $s_i^{(1)}=0,1$ and $s_i^{(2)}=0,1$, indicating the direction of particle-hop: 
If $s_i^{(1)}=1$ (or $s_i^{(2)}=1$), the first particle (or the second particle) goes to right, and $s_i^{(1)}, s_i^{(2)}=0$ otherwise. The transition rate from a configuration $\{m_i\}$ to another configuration $\{m_i'\}$ can be written as 
\be 
\Gamma [\{m_i\} \rightarrow \{m_i'\}] = \prod_i \phi(s_i^{(1)}, s_i^{(2)}),
\ee 
provided that the particular transition is allowed, i.e., the particular sites are active, etc. The {\it unbiased} dynamics with parallel update rules can be written as 
\bea
m_i(t+1) = [\hat a_i (m_i-2) + (1 - \hat a_i) m_i] 
\nonumber \\
+ \hat a_{i+1} [ (1-s_{i+1}^{(1)}) + (1-s_{i+1}^{(2)}) ] 
\nonumber \\ 
+ \hat a_{i-1} [s_{i-1}^{(1)} + s_{i-1}^{(2)}].
\eea
In the {\it biased} case, the energy cost $\Delta e_i$ associated with each of the toppling event at site $i$ can be written as
$$
\Delta e_i = \hat a_i [ (s_i^{(1)} + s_i^{(2)}) - \{ (1-s_i^{(1)}) + (1-s_i^{(2)}) \} ] F,
$$
which modifies the transition rate from a configuration $\{m_i\}$ to another configuration $\{m_i'\}$ as
$$
\Gamma [\{m_i\} \rightarrow \{m_i'\}] = \prod_i \frac{ \phi(s_i^{(1)}, s_i^{(2)}) e^{\Delta e_i/2} }{\gamma(F)},
$$
where the normalization factor $\gamma(F) = \sum_{s_i^{(1)}, s_i^{(2)}} \phi(s_i^{(1)}, s_i^{(2)}) e^{\Delta e_i/2} = 1 +{\cal O}(F^2) \approx 1$ as we only collect terms linear in $F$. 
Then the time evolution of first moment of mass with {\it biased} dynamics can be written as
\bea 
\langle m_i(t+1) \rangle = \langle [\hat a_i (m_i-2) + (1 - \hat a_i) m_i ] e^{\Delta e_i /2} \rangle 
\nonumber \\
+ \langle \hat a_{i+1} [(1-s_{i+1}^{(1)}) + (1-s_{i+1}^{(2)})] e^{\Delta e_{i+1}/2} \rangle 
\nonumber \\
+ \langle \hat a_{i-1} [s_{i-1}^{(1)} + s_{i-1}^{(2)}] e^{\Delta e_{i-1}/2} \rangle.
\nonumber
\eea
Expanding the rhs of the above equation in linear order of $F$, we obtain the time evolution equation for local density
$$
\rho_i (t+1) - \rho_i(t) = (a_{i-1} -2 a_i   + a_{i+1}) + F\frac{(a_{i-1} - a_{i+1})}{2}
$$ 
where $\langle \hat a_i \rangle = a_i$. In large spatio-temporal scales where observables are slowly varying function of space and time, local observable like activity $a_i[\rho_i(t)] \equiv a[\rho(x,t)]$, we obtain hydrodynamic evolution of density $\rho(x,t)$ at position $x$ and time $t$ in the CMS with parallel updates,
\be 
\frac{\partial \rho(x, t)}{\partial t} = \frac{\partial^2 a(\rho)}{\partial x^2} - F \frac{\partial a(\rho)}{\partial x}.
\ee
From the above equation, we get the expressions of diffusion current $J_{diff} \equiv - D(\rho) \partial \rho/\partial x$ and drift current $J_{drift} \equiv \chi(\rho) F$ where density-dependent bulk-diffusion coefficient and conductivity can be written as
$D(\rho) = {da}/{d\rho} \equiv a'(\rho)$ and $\chi(\rho) = a(\rho)$, respectively. As discussed before, following macroscopic fluctuation theory \cite{Bertini-PRL2001, Jona-Lasinio-RMP2015}, we recover the ER $\sigma^2(\rho) = {\chi(\rho)}/{D(\rho)} = {a(\rho)}/{a'(\rho)}$, or alternatively,
\begin{equation}
\sigma^2(\rho)  = \left[ \frac{d(\ln a)}{d \rho} \right]^{-1}.
\label{ER-FR-PU}
\end{equation}
In Fig. (\ref{ER-fig-DMS-PU}), we have plotted chemical potentials, $\mu(\rho) = \int_{\rho_0}^{\rho} 1/\sigma^2 d\rho$ obtained by integrating inverse of lhs of Eq. (\ref{ER-FR-PU}) and $\mu(\rho) = [\ln a(\rho) - \ln a(\rho_0)]$ obtained by integrating the inverse of the rhs of Eq. (\ref{ER-FR-PU}), as a function of density $\rho$. As seen in the plot, the two chemical potentials are in reasonably good agreement with each other, thus verifying the ER for the CMS with parallel update in one dimension.

\subsection{Continuous-mass stochastic sandpile}
\label{con-css}

In this section, we discuss the results in a continuous-mass version of the conserved stochastic sandpiles \cite{Mohanty-PRL2012}, which, for simplicity, we define on a periodic one-dimensional lattice of $L$ sites with $m_i \ge 0$ being a continuous (and unbounded) mass variable assigned to site $i$. Also, we consider here only the continuous-time dynamics (random sequential updates). The model can be straightforwardly generalized to higher dimensions and other update rules such as parallel updates. Total mass $M = \sum_{i=1}^L m_i$ remains {\it conserved} in the process, with mass-density $\rho=M/L$ fixed. Provided $m_i \ge 1$, site $i$ becomes {\it active} and topples with rate unity, by transferring a uniformly distributed random fraction $\xi_i \in [0,1]$ of mass $m_i$ to its left nearest neighbor and the rest of the mass to its right nearest neighbor. The system undergoes an active to absorbing phase transition below a critical density $\rho_c \approx 0.66$.

To calculate the conductivity, we now bias the system by applying a small constant force $\vec{F}=F \hat{x}$, which leads to the following evolution of mass $m_i(t)$ in the infinitesimal time interval $dt$,
\begin{eqnarray}
m_i(t+dt) =
\left\{
\begin{array}{ll} 
m_i(t) - \xi_i m_i(t)  & {\rm prob.}~\hat a_i c_i^Fdt, \cr
m_i(t) + \xi_{i+1} m_{i+1}(t)  & {\rm prob.}~\hat a_{i+1} c_{i+1}^{F}dt, \cr 
m_i(t) + \tilde \xi_{i-1} m_{i-1}(t) & {\rm prob.}~\hat a_{i-1} c_{i-1}^{F}dt, \cr
m_i(t) & {\rm otherwise},
\end{array}
\right.
\label{continuous_manna-fes-dynamics_biased}
\end{eqnarray}
where the random variable $\hat a_i = 1$ if a site is active and $\hat a_i=0$ otherwise, $\tilde \xi_i = (1 - \xi_i)$, modified (biased) mass-transfer rates $c_{i}^F = c^{F=0}_{i} \exp [ \sum_j \Delta e_{ij}/2 ]$ from site $i$ to nearest neighbors, with the corresponding unbiased ($F=0$) mass-transfer rates $c_i^{F=0} = 1$, $\Delta e_{ij} = \Delta m_{i \rightarrow j} F (j-i)b$ being an `energy cost' to transfer $\Delta m_{i \rightarrow j}$ amount of mass from site $i$ to $j$ (lattice spacing $b=1$). Now, by keeping only the terms linear in biasing force $F$,
\begin{eqnarray}
c_i^F = \hat a_i \left[ 1 + \frac{m_i (1 - 2 \xi_i)}{2} F \right],
\end{eqnarray} 
we arrive at the time-evolution equation of local mass-density $\langle m_i(t) \rangle = \rho_i(t)$, 
\begin{eqnarray}
\frac{\partial \rho_i}{\partial t} = \left[ u^{(1)}_{i-1} - 2 u^{(1)}_i + u^{(1)}_{i+1} \right] + F \left[ u^{(2)}_{i-1} - u^{(2)}_{i+1} \right],
\label{GP-CMS}
\end{eqnarray}
where we denote $u^{(1)}_i = \langle m_i \hat a_i \rangle/2$ and $u^{(2)}_i = \langle m_i^2 \hat a_i \rangle/12$. Note that the gradient property is still satisfied by the time-evolution equation as the rhs can be written as gradients (discrete) of the two local observables $u^{(1)}$ and $u^{(2)}$. Now, in large spatio-temporal scales where observables are slowly varying functions of space and time, local observables $u^{(\alpha)} = u^{(\alpha)}[\rho_i(t)] \equiv u^{(\alpha)}[\rho(x,t)]$, with $\alpha=1,2$, are functions of only local density $\rho(x,t)$. Therefore, using the property of local steady state and in the continuum limit, we obtain, from Eq. \ref{GP-CMS}, the hydrodynamic evolution of density field $\rho(x,t)$ at position $x$ and time $t$, 
\begin{eqnarray}
\frac{\partial \rho(x,t)}{\partial t} = \frac{\partial^2 u^{(1)}}{\partial x^2} - \frac{\partial u^{(2)}}{\partial x} \equiv - \frac{\partial J}{\partial x}.
\label{Hyd-CMS}
\end{eqnarray}
In the above equation, the local current $J(\rho(x)) = J_{diff} + J_{drift}$ can be decomposed into two parts: Diffusive current $$J_{diff} = - \frac{\partial u^{(1)}}{\partial x},$$ and drift current $$J_{drift} = u^{(2)} F,$$ leading to the expressions for bulk-diffusion coefficient $D(\rho) = du^{(1)}/d\rho$ and conductivity $\chi(\rho) = u^{(2)}$, both of which are density-dependent. Then, by using macroscopic fluctuation theory \cite{Bertini-PRL2001, Jona-Lasinio-RMP2015}, via Eqs. (\ref{additivity}) and (\ref{MFT}), we obtain an Einstein relation (ER) as in Eq. (1), between scaled variance $\sigma^2(\rho)$ of subsystem mass, bulk-diffusion coefficient $D(\rho)$ and conductivity $\chi(\rho)$.

\begin{figure}
\includegraphics[scale=0.7]{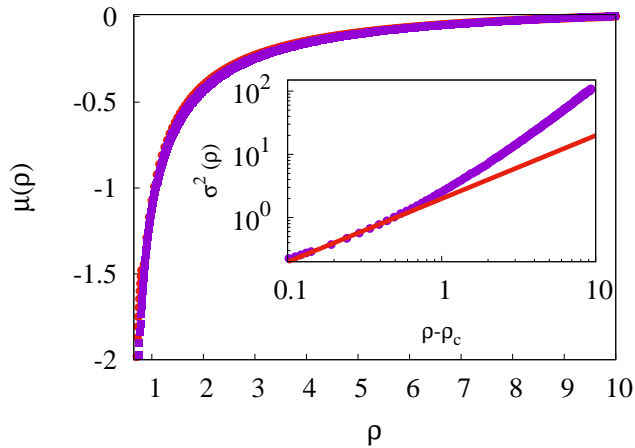}
\caption{Verification of Einstein relation in the continuous-mass CSS: Chemical potentials $\mu(\rho) = \int_{\rho_0}^{\rho} D(\rho)/\chi(\rho) d\rho$, with $D(\rho) = d u^{(1)}/d\rho$ and $\chi(\rho)=u^{(2)}(\rho)$, and $\mu(\rho) = \int_{\rho_0}^{\rho} 1/\sigma^2(\rho) d\rho$ are plotted as function of density $\rho$ [obtained by integrating the inverse of the rhs and lhs of Eq. (\ref{ER})]. {\it Inset}: The scaled variance $\sigma^2(\rho)$ is plotted as a function $(\rho - \rho_c)$ with $\rho_c \approx 0.66$, demonstrating $\sigma^2(\rho) \propto (\rho-\rho_c)$. Points - simulations, red line - theory [scaling relation (ii) without theoretical determination of the proportionality constant].  
}
\label{ER-fig-CMS}
\end{figure}

In Fig. \ref{ER-fig-CMS}, we have plotted chemical potentials $\mu(\rho) = \int_{\rho_0}^{\rho} (1/\sigma^2) d\rho$ and $\mu(\rho) = \int_{\rho_0}^{\rho} (D/\chi) d\rho$, obtained by numerically (simulations) calculating $u^{(1)}(\rho)$ [thus also $du^{(1)}/d\rho$) and $u^{(2)}(\rho)$] as a function of $\rho$ and then numerically integrating the inverse of Eq. (1). Both chemical potentials are in reasonably good agreement with each other, thus verifying the ER Eq. (\ref{ER}) in the continuous-mass CSS.

Now, assuming that, near criticality, singularities in the quantities $u^{(1)}(\rho) = (1/L) \sum_i \langle m_i \hat a_i \rangle/2$ and $u^{(2)}(\rho) = (1/L)\sum_i \langle m_i^2 \hat a_i \rangle/12$ come from the singular contribution of only the activity $a(\rho) \sim (\rho-\rho_c)^{\beta}$, i.e., as $\rho \rightarrow 0^+$,
$$
u^{(1)}(\rho) \simeq {\rm const.} a(\rho)~;~u^{(2)}(\rho) \simeq {\rm const.} a(\rho),
$$ 
we recover scaling relations (i) $\sigma^2(\rho) \propto (\rho-\rho_c)$ and (ii) $z=2+(\beta-1)/\nu_{\perp}$ as in the case of the conserved Manna sandpiles. Here we use the ER Eq. (\ref{ER}) and the fluctuation-response relation Eq. (\ref{FR}) to have bulk-diffusion coefficient $D(\rho) \sim a'(\rho)$, conductivity $\chi(\rho) \sim a(\rho)$, and consequently chemical potential $\mu(\rho) = \int 1/\sigma^2 d\rho = \int D(\rho)/\chi(\rho) d\rho \sim \ln a(\rho)$. The above analysis is however valid only near criticality. Note that, in the case of continuous-mass CSS, the proportionality constant in scaling relation (i) however could not be determined. In the inset of Fig. \ref{ER-fig-CMS}, we plot scaled variance $\sigma^2(\rho)$ of subsystem mass as a function of $(\rho-\rho_c)$, which is in quite good agreement with simulations.

\section{Summary and conclusions}
\label{Summary}

We derive an exact hydrodynamic structure of a broad class of conserved-mass ({\it fixed-energy}) stochastic sandpiles (CSS). Importantly, these systems possess a `gradient property', where local diffusive current and, therefore, time-evolution of local densities [see the rhs of Eq. (\ref{GP-DMS})], can be written as a gradient (discrete) of local observable like the activity. The gradient property essentially originates from the fact that, in the sandpiles studied here, the particle hopping rates depend only on the departure site, but not on the destination sites. We use the property, and recently developed macroscopic fluctuation theory \cite{Bertini-PRL2001, Jona-Lasinio-RMP2015}, to uncover a remarkable thermodynamic structure, where bulk-diffusion coefficient $D(\rho)$, conductivity $\chi(\rho)$ and mass fluctuations $\sigma^2(\rho)$ are shown to be connected to the activity $a(\rho)$, through an equilibrium-like Einstein relation $\sigma^2(\rho) = \chi(\rho)/D(\rho)$.

In particular, in the conserved Manna sandpiles (CMS), we have strikingly simple relations $D(\rho)=a'(\rho)$, $\chi(\rho) = a(\rho)$ and therefore $$\sigma^2(\rho) = \frac{a(\rho)}{a'(\rho)}.$$ Moreover, in the CMS, we compute probabilities of density large deviations, which is governed by an equilibrium-like chemical potential $\mu(\rho)$ directly related again to the activity $a(\rho)$ as $$\mu(\rho) = \ln a(\rho) + {\rm const.}$$ 
Our theoretical results, predicted by the hydrodynamics as described in Eq. (\ref{Hyd-biased}), are in quite good agreement with simulations and have been generalized to higher dimensions, other update rules (e.g., parallel update), and a continuous-mass version of the CSS.

The hydrodynamic structure obtained here has far-reaching consequences on the critical behaviors of the CSSs, through the two scaling relations (i) and (ii), which, we believe, could help settle the long-standing issue of universality in such systems. As evident from Table 1, a broad class of the CSSs - restricted-height versions \cite{Dickman-PRE2006}, conserved lattice gases (CLGs) \cite{Rossi-PRL2000} and conserved threshold transfer processes (CTTPs) \cite{Lubeck-PRE} - all obey reasonably well scaling relation (ii). The scaling relation is manifestly violated for directed percolation (DP), which does not  have any conservation law as such, thus ruling out DP universality for the conserved Manna sandpiles (CMSs) in particular and, presumably, for the CSSs in general. However, unlike the CMSs, many of the CSSs, with bounded state-space, can have `non-gradient' structures in density evolution, where local current cannot be written as a gradient of a local variable. The issue of putting the latter assertion regarding universality in the CSS on a firmer ground requires further studies and remains open, and intriguingly poised.

Interestingly, assuming the equivalence between the CSS and the quenched Edwards-Wilkinson (qEW) interface model -  a long-standing conjecture \cite{Vespignani-PRE2000, Grassberger-PRE2016, LeDoussal-PRL2015, Paczuski} - and that between the SOC sandpiles and the conserved sandpiles at criticality, one might argue in support of scaling relation (ii) (see \cite{note}). However, it should also be noted that no rigorous connection has so far been established between the CSS and the interface models \cite{Vespignani-PRE2000, Grassberger-PRE2016, LeDoussal-PRL2015}, or between the SOC version and conserved version of sandpiles.  In this scenario, our hydrodynamic theory could provide useful insights into the above possible connections, especially the latter one where the hydrodynamic description could be applicable with appropriate open boundary conditions.

\begin{table}
\caption{Previous estimates of $z$ are compared with $z$, calculated from scaling relation (ii) [using previously estimated static exponents $\beta$ and $\nu_{\perp}$ in (ii)].}
\begin{center}
  \begin{tabular}{ | p{3.5cm} | p{1.1cm} | p{1.1cm} | p{1.1cm} | p{1.4cm} | } \hline
  
     Models: Conserved stochastic sandpiles (CSS) and directed percolation (DP) & $\beta$ & $\nu_{\perp}$ & $z$ & $z$ from scaling relation (ii) \\ \hline 
     
    1D {\it unrestricted-height} MS, from Ref. \cite{Dickman-PRE2001} & $0.42$ &  $1.81$ & $1.66$ & $1.68$  \\ \hline
    
    2D {\it unrestricted-height} MS, from Ref. \cite{Vespignani-PRE2000} & 0.64 & 0.82 & 1.57 & 1.56 \\ \hline
    
    1D {\it restricted-height} MS, from Ref. \cite{Dickman-PRE2006} & 0.29 & 1.36 & 1.50 & 1.48 \\ \hline
    
    2D {\it restricted-height} MS, from Ref. \cite{Dickman-JSTAT2014} & 0.64 & 0.82 & 1.51 & 1.56 \\ \hline
    
    1D conserved lattice gas (CLG), from Ref. \cite{Rossi-PRL2000} & 0.63 & 0.78 & 1.52 & 1.53 \\ \hline 
    
    2D conserved threshold transfer process (CTTP), from Table 1 in Ref. \cite{Dickman-JSTAT2014} & 0.64 & 0.80 & 1.53 & 1.55 \\ \hline
    \hline
    
    1D DP, from Table 1 in Ref. \cite{Munoz-PRE2004} & 0.28 & 1.10 & 1.58 & 1.35 \\ \hline
    
    2D DP, from Table 1 in Ref. \cite{Dickman-JSTAT2014} & 0.58 & 0.73 & 1.77 & 1.42 \\ \hline
     
\end{tabular}
 
\end{center}
\label{expo-table}
\end{table}


\section{Acknowledgement} 

We thank Deepak Dhar for careful reading of the manuscript and for invaluable comments and suggestions, and thank one of the anonymous referees for suggesting the point discussed in \cite{note}. P.P. thanks Pradeep Mohanty for general discussions on active-absorbing phase transitions and the corresponding literature. P.P. gratefully acknowledges financial support from the Science and Engineering Research Board (SERB), India [Grant No. EMR/2014/000719]. S.C. acknowledges financial support from the Council of Scientific and Industrial Research (CSIR), India [Grant No. 09/575(0099)/2012-EMR-I] for part of the work carried out under her senior research fellowship. S.C. and A.D. gratefully acknowledge extended research fellowship, provided by the S. N. Bose National Centre for the Basic Sciences, Kolkata, India.


\begin{thebibliography}{99}

\bibitem{BTW} P. Bak, C. Tang, and K. Wiesenfeld,
Phys. Rev. Lett. {\bf 59}, 381 (1987); {\it ibid}, Phys. Rev. A {\bf 38}, 364 (1988).

\bibitem{Kadanoff} L. P. Kadanoff, S. R. Nagel, L. Wu, and S. Zhou, Phys. Rev. A {\bf 39}, 6524 (1989).

\bibitem{SOC} P. Bak, {\it How Nature Works: The Science of Self-Organised Criticality} (Copernicus, New York, 1996).

\bibitem{Kardar-Nature1996}  Sidney R. Nagel, Rev. Mod. Phys. {\bf 64}, 321 (1992); Kardar, Nature (London) 379, {\bf 22}  (1996).

\bibitem{Hazra} A. Cipriani, R. S. Hazra, and W. M. Ruszel,  	{\it arXiv:1610.09863}; L. Levine, M. Murugan, Y. Peres, and B. E. Ugurcan, Ann. Henri Poincare, 1 (2015); L. Levine and Y. Peres, J. Anal. Math. {\bf 111}, 151 (2010).

\bibitem{Dhar-exact} D. Dhar and R. Ramaswamy, Phys. Rev. Lett. {\bf 63}, 1659 (1989); D. Dhar, Phys. Rev. Lett. {\bf 64}, 1613 (1990); V. B. Priezzhev, D. Dhar, A. Dhar, and S. Krishnamurthy, Phys. Rev. Lett. {\bf 77}, 5079 (1996).

\bibitem{Priezzhev-exact} V. B. Priezzhev, D. V. Ktitarev, and E. V. Ivashkevich, Phys. Rev. Lett. {\bf 76}, 2093 (1996); V. B. Priezzhev, J. Stat. Phys. {\bf 74}, 955 (1994); E. V. Ivashkevich, J. Phys. A {\bf 27}, 3643 (1994). D. V. Ktitarev, S. Lubeck, P. Grassberger, and V. B. Priezzhev, Phys. Rev. E {\bf 61}, 81 (2000).

\bibitem{Manna} S. S. Manna, J. Phys. A {\bf 24}, L363 (1991); {\it ibid}, J. Stat. Phys. {\bf 59}, 509 (1990).

\bibitem{Vespignani-PRE2000} A. Vespignani, R. Dickman, M. A. Munoz, and S. Zapperi, Phys. Rev. E {\bf 62}, 4564 (2000).

\bibitem{Rossi-PRL2000} M. Rossi, R. Pastor-Satorras, and A. Vespignani, Phys. Rev. Lett. {\bf 85}, 1803 (2000).

\bibitem{Dickman-PRE2001} R. Dickman, M. Alava, M. A. Munoz, J. Peltola, A. Vespignani, and S. Zapperi, Phys. Rev. E {\bf 64}, 056104 (2001).

\bibitem{Dickman-PRE2002} R. Dickman, T. Tome and M. J. de Oliveira, Phys. Rev. E {\bf 66}, 016111 (2002).

\bibitem{Lubeck-PRE} S. Lubeck, Phys. Rev. E {\bf 66}, 046114 (2002); S. Lubeck and P. C. Heger, Phys. Rev. E {\bf 68}, 056102 (2003); S. Lubeck and P. C. Heger, Phys. Rev. Lett. {\bf 90}, 230601 (2003).

\bibitem{Bonachela-PRE2008} J. A. Bonachela and M. A. Munoz, Phys. Rev. E {\bf 78}, 041102 (2008); G. Menon and R. Ramaswamy, Phys. Rev. E {\bf 79}, 061108 (2009).

\bibitem{Dickman-JSTAT2014} S. D. da Cunha, L. R. da Silva, G. M. Viswanathan, and R. Dickman, J. Stat. Mech. {\bf P08003} (2014).

\bibitem{Hexner-PRL2015} D. Hexner and D. Levine, Phys. Rev. Lett. {\bf 114}, 110602 (2015); {\it ibid}, Phys. Rev. Lett. {\bf 118}, 020601 (2017). 

\bibitem{Grassberger-PRE2016} P. Grassberger, D. Dhar, and P. K. Mohanty, Phys. Rev. E {\bf 94}, 042314 (2016).

\bibitem{Nagel} H. M. Jaeger, C. Liu, and S. R. Nagel, Phys. Rev. Lett. {\bf 62}, 40 (1989).

\bibitem{Oslo} V. Frette, K. Christensen, A. Malthe-Sorenssen, J. Feder, T. Jossang and P. Meakin, Nature (London) 379, {\bf 49} (1996).

\bibitem{review} R. Dickman, M. A. Munoz, A. Vespignani, and S. Zapperi, Braz. J. Phys. {\bf 30}, 27 (2000); D. Dhar, Phys. A (Amsterdam) {\bf 369}, 29 (2006). 

\bibitem{Vespignani-PRL1998} A. Vespignani, R. Dickman, M. A. Munoz, and S. Zapperi, Phys. Rev. Lett. {\bf 81}, 5676 (1998); R. Dickman, A. Vespignani, and S. Zapperi, Phys. Rev. E {\bf 57}, 5095 (1998).

\bibitem{APT} J. Marro and R. Dickman, {\it Nonequilibrium Phase Transitions in Lattice Models} (Cambridge University Press,
Cambridge, United Kingdom, 1999).

\bibitem{Hwa-PRL1989} T. Hwa and M. Kardar, Phys. Rev. Lett. {\bf 62}, 1813 (1989).

\bibitem{Munoz-PRL1996} M. A. Munoz, G. Grinstein, R. Dickman, and R. Livi, Phys. Rev. Lett. {\bf 76}, 451 (1996). 

\bibitem{Munoz-PRE2004} J. J. Ramasco, M. A. Munoz,  and C. A. da Silva Santos, Phys. Rev. E {\bf 69}, 045105(R) (2004).

\bibitem{Munoz-PRL2005} I. Dornic, H. Chate and M. A. Munoz, Phys. Rev. Lett. {\bf 94}, 100601 (2005).

\bibitem{Carlson} J. M. Carlson, J. T. Chayes, E. R. Grannan, and G. H. Swindle, Phys. Rev. Lett. {\bf 65}, 2547 (1990).

\bibitem{Pradhan-JSTAT2004} D. Dhar and P. Pradhan, J. Stat. Mech.: Theory Exp. {\bf P05002} (2004).

\bibitem{Pradhan-PRE2006} P. Pradhan and D. Dhar, Phys. Rev. E {\bf 73}, 021303 (2006).

\bibitem{Dickman-EPJB2009} S. D. da Cunha, R. R. Vidigal, L. R. da Silva and R. Dickman, Eur. Phys. B {\bf 72}, 441 (2009).

\bibitem{Dickman-PRE2006} R. Dickman, Phys. Rev. E {\bf 73}, 036131 (2006).

\bibitem{Munoz-PRL2007} J. A. Bonachela, H. Chate, I. Dornic, and M. A. Munoz, Phys. Rev. Lett. {\bf 98}, 155702 (2007).

\bibitem{Mohanty-PRL2002} P. K. Mohanty and D. Dhar, Phys. Rev. Lett. {\bf 89}, 104303 (2002).

\bibitem{Mohanty-PRL2012} M. Basu, U. Basu, S. Bondyopadhyay, P. K. Mohanty, and H. Hinrichsen, Phys. Rev. Lett. {\bf 109}, 015702 (2012). 

\bibitem{Lee} S. B. Lee, Phys. Rev. Lett. {\bf 110}, 159601 (2013);  {\it ibid}, Phys. Rev. E {\bf 89}, 062133 (2014); {\it ibid}, Phys. Rev. E {\bf 89} 060101 (2014).

\bibitem{Dickman-PRE2015}  R. Dickman and S. D. da Cunha, Phys. Rev. E {\bf 92}, 020104(R) (2015).

\bibitem{LeDoussal-PRL2015} P. Le Doussal and K. J. Wiese, Phys. Rev. Lett. {\bf 114}, 110601 (2015).

\bibitem{Biham-PRE1996} A. Ben-Hur and O. Biham, Phys. Rev. E {\bf 53}, R1317 (1996).

\bibitem{Grassberger-JSP1995} P. Grassberger, J. Stat. Phys. {\bf 79}, 13 (1995).

\bibitem{DP} M. Henkel, H. Hinrichsen, and S. Lubeck, {\it Non-Equilibrium Phase Transitions Vol. I: Absorbing Phase
Transitions} (Dordrecht, Springer, 2008).

\bibitem{Grassberger-1981} P. Grassberger, Z. Phys. B {\bf 42}, 151 (1981).


\bibitem{Eyink} G. Eyink, J. L. Lebowitz and H. Spohn,
Comm. Math. Phys. {\bf 132}, 253 (1990); {\it ibid}, {\bf 140}, 119 (1991).

\bibitem{Carlson2} J. M. Carlson, E. R. Grannan, G. H. Swindle, and J. Tour, Ann. Prob. {\bf 21}, 1372 (1993).

\bibitem{Landim} C. Kipnis and C. Landim, {\it Scaling Limits of Interacting Particle Systems} (Berlin: Springer, 1999); M. Z. Guo, G. C. Papanicoiaou, S. R. S. Varadhan, Comm. Math. Phys. {\bf 118}, 31 (1988).


\bibitem{Bertini-PRL2001} L. Bertini, A. D. Sole, D. Gabrielli, G. Jona-Lasinio, and C. Landim, Phys. Rev. Lett. {\bf 87}, 040601 (2001);  {\it ibid}, J. Stat. Phys. {\bf 107}, 635 (2002).

\bibitem{Jona-Lasinio-RMP2015} L. Bertini, A. De Sole,
D. Gabrielli, G. Jona-Lasinio, and C. Landim, Rev. Mod. Phys.
{\bf 87}, 593 (2015).

\bibitem{Das-PRE2017} A. Das, A. Kundu, and P. Pradhan, Phys. Rev. E {\bf 95}, 062128 (2017).

\bibitem{Chatterjee-PRL2014} S. Chatterjee, P. Pradhan, and P. K. Mohanty, Phys. Rev. Lett. {\bf 112}, 030601 (2014).

\bibitem{Chatterjee-PRE2016} A. Das, S. Chatterjee, and P. Pradhan, Phys. Rev. E {\bf 93}, 062135 (2016).

\bibitem{Halperin-Hohenberg} P. C. Hohenberg and B. J. Halperin, Rev. Mod. Phys. {\bf 49}, 435 (1977).

\bibitem{Subhadip} Private communications with Subhadip Chakraborti.

\bibitem{Paczuski} M. Paczuski and S. Boettcher, Phys. Rev. Lett. {\bf 77}, 111 (1996).

\bibitem{note} Using renormalization group method in the qEW equation, one gets $\nu_{\perp} = 1/(2- \alpha)$ \cite{Barabasi}. In $d$ dimensions, the roughness exponent in the corresponding interface model $\alpha = D - d$ can be related to avalanche dimension $D$ of the SOC sandpile (slowly driven, dissipative) \cite{Paczuski} and the avalanche dimension $D = z + d - \beta/\nu_{\perp}$ to dynamic exponent $z$ in sandpile \cite{Pruessner}, leading to our scaling relation (ii). 

\bibitem{Barabasi} {\it Fractal Concepts in Surface Growth},  A.-L. Barabasi and H. E. Stanley, p. 110 (Cambridge University Press, 1995).

\bibitem{Pruessner} {\it Self-organized Criticality, Theory, Models, and Characterisation}, G. Pruessner (Cambridge University Press, 2012).


\end{thebibliography}
\end{document}